\documentclass[letter,traditabstract]{aa} 
\usepackage{graphicx}
\usepackage{natbib}
\usepackage{epsfig}
\usepackage{txfonts}
\usepackage{amssymb}
\usepackage{hyperref}

\newcommand{\ltsim}{\protect\raisebox{-0.5ex}{$\:\stackrel{\textstyle <}
        {\sim}\:$}}
\newcommand{\gtsim}{\protect\raisebox{-0.5ex}{$\:\stackrel{\textstyle >}
        {\sim}\:$}}

\bibpunct{(}{)}{;}{a}{}{,}

\def\teff{$T_\mathrm{eff}$}
\def\vsini{$v$\,sin\,$i$}                
\def\ms{\hbox{\,m\,s$^{-1}$}}            
\def\m2s2{\hbox{\,m$^{2}$\,s$^{-2}$}}    
\def\kms{\hbox{\,km\,s$^{-1}$}}          
\def\gcm3{\hbox{\,g\,cm$^{-3}$}}         
\def\vsini{\hbox{$v$\,sin\,$i_{\star}$}} 
\def\Msun{\hbox{$M_{\odot}$}}            
\def\Rsun{\hbox{$R_{\odot}$}}            
\def\Mjup{\hbox{$\mathrm{M}_{\rm Jup}$}} 
\def\Rjup{\hbox{$\mathrm{R}_{\rm Jup}$}} 
\def\degr{\hbox{$^\circ$}}               

\begin{document}
   \title{Doppler tomography of transiting exoplanets:\\
   A prograde, low-inclined orbit for the hot Jupiter 
   CoRoT-11b}


   \author{D.~Gandolfi\inst{\ref{ESA}}
      \and A.~Collier Cameron\inst{\ref{SUPA}}
      \and M.~Endl\inst{\ref{McD}}
      \and A.\,F.~Lanza\inst{\ref{OACT}}
      \and C.~Damiani\inst{\ref{OACT},\ref{LAM}}
      \and R.~Alonso\inst{\ref{IAC},\ref{LaLaguna}}
      \and W.\,D.~Cochran\inst{\ref{McD}}      
      \and M.~Deleuil\inst{\ref{LAM}}
      \and M.~Fridlund\inst{\ref{ESA}}
      \and A.\,P.~Hatzes\inst{\ref{Tautenburg}}
      \and E.\,W.~Guenther\inst{\ref{Tautenburg}}
          }

\institute{Research and Scientific Support Department, ESTEC/ESA, PO Box 299, 2200 AG Noordwijk, The Netherlands\label{ESA}\\ 
           \email{davide.gandolfi@esa.int}
      \and School of Physics \& Astronomy, University of St. Andrews, North Haugh, St. Andrews, Fife KY16 9SS, United Kingdom,\label{SUPA}
      \and McDonald Observatory, University of Texas at Austin, Austin, TX 78712, USA\label{McD}
      \and INAF - Osservatorio Astrofisico di Catania, Via S. Sofia, 78, 95123 Catania, Italy\label{OACT}
      \and Laboratoire d'Astrophysique de Marseille, CNRS \& University of Provence, 38 rue Fr\'ed\'eric Joliot-Curie, 13388 Marseille cedex 13, France\label{LAM}
      \and Instituto de Astrof\'\i sica de Canarias, C/\,V\'\i a L\'actea s/n, 38205 La Laguna, Spain\label{IAC}
      \and Departamento de Astrof\'\i sica, Universidad de La Laguna, 38206 La Laguna, Spain\label{LaLaguna}
      \and Th\"uringer Landessternwarte, Sternwarte 5, Tautenburg, D-07778 Tautenburg, Germany\label{Tautenburg}
           }

\date{Received 3 May 2012; accepted 12 June 2012}

 
\abstract{We report the detection of the Doppler shadow of the transiting hot Jupiter CoRoT-11b. 
          Our analysis is based on line-profile tomography of time-series, Keck/HIRES 
          high-resolution spectra acquired during the transit of the planet. We measured 
          a sky-projected, spin-orbit angle $\lambda=0.1\degr\pm2.6\degr$, which is consistent with 
          a very low-inclined orbit with respect to the stellar rotation axis. We refined the physical 
          parameters of the system using a Markov chain Monte Carlo simultaneous fitting of the available 
          photometric and spectroscopic data. An analysis of the tidal evolution of the system shows how 
          the currently measured obliquity and its uncertainty translate into an initial absolute value 
          of less than about $10\degr$ on the zero-age main sequence, for an expected average modified 
          tidal quality factor of the star $\langle Q\,_{\star}^{\prime} \rangle \gtsim 4 \times 10^{6}.$
          This is indicative of an inward migration scenario that would not have perturbed the primordial 
          low obliquity of CoRoT-11b. Taking into account the effective temperature and mass of the planet 
          host star (\teff\,=\,6440\,K, $M_\star$\,=\,1.23\,\Msun), the system can be considered a new 
          telling exception to the recently proposed trend, according to which relatively hot and massive 
          stars (\teff\,$>$\,6250\,K, $M_\star$\,$>$\,1.2\,\Msun) seem to be preferentially orbited by 
          hot Jupiters with high obliquity.}

\keywords{planetary systems - planet-star interaction - stars: individual: \object{CoRoT-11}}
               
\titlerunning{A prograde, low-inclined orbit for the hot Jupiter CoRoT-11b}
\authorrunning{Gandolfi et al.}

   \maketitle
%

\section{Introduction}
\label{Introduction}

Time-series, high-resolution spectroscopic observations of planetary 
transits allow us to detect the Rossiter-McLaughlin (RM) effect 
\citep{Rossiter1924,McLaughlin1924} and to measure the sky-projected 
system obliquity $\lambda$, i.e., the angle in the plane of the sky 
between the projections of the planet's orbital angular momentum and 
the star's rotation spin. When a planet transits in front of its host 
star, it subsequently occults different parts of the stellar disc. 
For a rotating star, this results in a distortion of the stellar 
line profiles that change during the transit. If stellar rotation is 
the dominant source of line-broadening, high-resolution spectroscopy 
can resolve the stellar line profiles and shows the distortion (bump) 
moving across the line profiles, which is the result of the missing 
starlight blocked by the transiting planet \citep[e.g.,][]{Cameron2010a,
Cameron2010b,Miller2010}. On the other hand, if stellar rotation is 
not the dominant broadening mechanism, the RM effect is detected as 
an anomalous Doppler shift of the stellar lines \citep[e.g.,][]
{Queloz2000,Ohta2005,Gimenez2006,Gaudi2007,Winn2010a}.

Giant planets with a semi-major axis $a\lesssim0.1$~AU (hot Jupiters)
are thought to have formed at large orbital distances from their 
parent star and then migrated inwards to their current position 
through a) tidal interaction with the proto-planetary disc in the 
pre-main sequence phase of the star/planet evolution \citep[e.g.,]
[]{Lin1996}; b) planet-planet scattering \citep[e.g.,][]{Rasio1996}; 
c) gravitational interaction with a third outer planet or stellar 
companion \citep[Kozai mechanism; e.g.][]{Kozai1962, Fabrycky2007,
Wu2007,Naoz2011}. The migration mechanism is not completely understood 
yet and the proposed theories predict different final obliquities with 
low values in the case of a migration governed by planet-disc interaction 
and significant misalignment in the case of the other mechanisms 
\citep{Triaud2010, Morton2011}. Measuring the spin-orbit obliquity 
of hot Jupiters can thus provide insights into their migration 
mechanism and help to discern between the rivaling theories.

\citet{Winn2010b} empirically found that hot Jupiters orbiting stars 
with an effective temperature \teff\,$\ge6250$ \,K\ and a mass 
$M_\star\ge1.2$\,\Msun\ tend to be misaligned. A similar result was 
obtained by \citet{Schlaufman2010}, who found evidence of spin-orbit 
misalignment along the line of sight in systems with massive stars and 
planets. According to \citet{Winn2010b}, either a planet formation 
scenario dependent on the stellar mass, or an efficient tidal realignment 
of hot Jupiters around cool stars, might account for the suspected trend. 
More recently, \citet{Triaud2011} suggested that the relatively rapid 
evolution of stars with $M_\star\ge1.2$\,\Msun\ combined with an 
observational bias, might explain the lack of aligned systems for 
stars with \teff\,$\ge6250$\,K. However, the number of exoplanets with 
secure measurements of the $\lambda$ angle is not yet statistically 
significant enough to confirm this trend, as pointed out by \citet{Moutou2011}.  

\citet{Gandolfi2010} announced the discovery of the transiting hot Jupiter 
CoRoT-11b, a 2.33~\Mjup\ planet in a 2.99 days orbit around a \teff=6440\,K 
dwarf star (SpT=F6\,V) with a relatively high projected rotational velocity 
(\vsini\,=\,40~\kms). With the aim of detecting the RM effect of the planet, 
the authors acquired radial velocity (RV) measurements during the transit, 
using the High Resolution Echelle Spectrometer (HIRES) mounted on the Keck I 
10\,m telescope, at the Keck Observatory (Mauna Kea, Hawai'i). Unfortunately, 
the observations were scheduled according to an old, slightly incorrect transit 
ephemeris, resulting in a partial coverage of the event. Although the HIRES 
RV data clearly show the RM anomaly of CoRoT-11, \citet{Gandolfi2010} were 
unable to place strong constraints on the orbit obliquity, owing to the partial 
coverage of the transit event and the relatively high \vsini\ of the star, 
which strongly affected the precision of their RV measurements.

In this letter, we show how the problems encountered in deriving 
accurate RV measurements of rapidly rotating stars to assess the shape of 
their RM anomaly can be overcome with Doppler imaging techniques, even in 
the occurrence of a partial spectroscopic coverage of the transit event. 
On the basis of line-profile tomography applied to the above-mentioned HIRES 
spectra, we succeed in detecting the Doppler shadow of CoRoT-11b and 
measuring the sky-projected obliquity of the system. A simultaneous 
reanalysis of the available photometric and spectroscopic data using 
a Markov chain Monte Carlo (MCMC) approach, enabled us to refine 
the system's parameters.


\section{Observations and data analysis}

As part of NASA's key science project to support the CoRoT mission, 
high-resolution ($R\approx 50\,000$) transit spectroscopy was 
carried out using the HIRES spectrograph \citep{Vogt94}. A detailed 
description of the adopted instrument set-up and data reduction 
was reported in \citet{Gandolfi2010}. Twelve consecutive spectra 
of 900 seconds each were gathered during the transit that occurred 
on the night of 1 July 2009. The observations started about 
1.2 hours before the beginning of the transit and erroneously 
stopped $\sim$45 minutes before the end of the event, owing to an 
incorrect transit ephemeris at the time of the observations. The 
altitude of CoRoT-11 increased during the time-series observations, 
with an airmass varying from 1.80 to 1.05. Two more out-of-transit 
spectra were secured on the night before and one on the same night, 
to properly combine the HIRES RV measurements with data previously 
taken with other instruments.

The HIRES spectra were acquired using an $I_2$ absorption cell 
to track any instrumental drifts and improve the RV precision. 
Since our analysis is based on Doppler imaging of the stellar 
line profile, we used only the echelle orders that were not 
contaminated by the $I_2$ absorption lines. The first spectrum
was discarded as affected by the high airmass of the exposure. 
This resulted in a set of eleven consecutive HIRES spectra covering 
the wavelength range $3800-4860$\,\AA, with a signal-to-noise ratio 
ranging from 40 to 70 per pixel at 4300\,\AA.

\begin{figure}
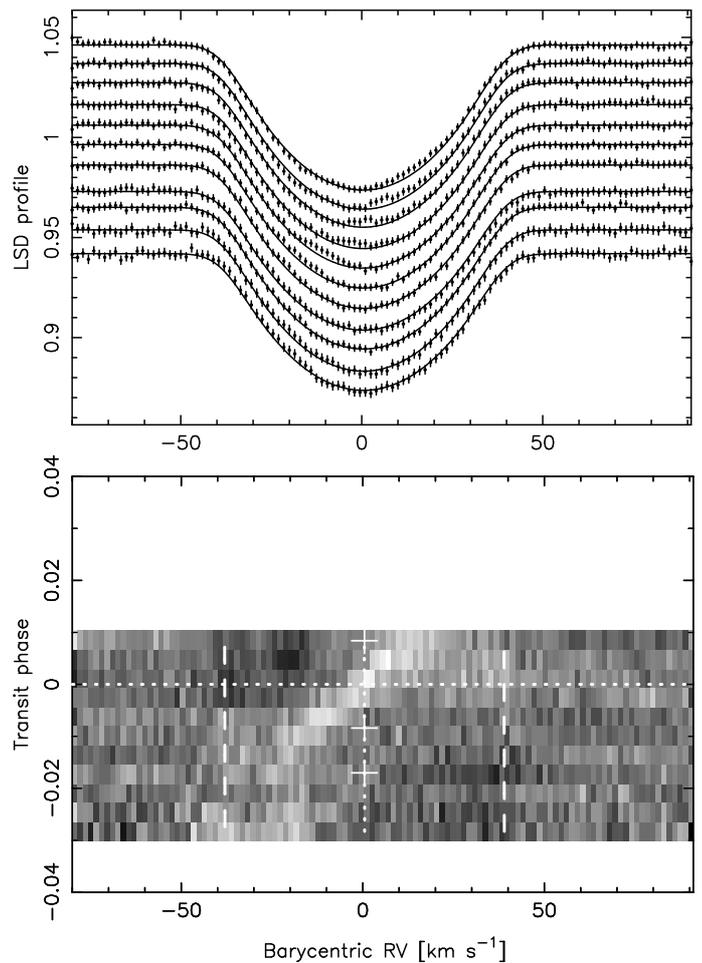
 
\begin{center}
\resizebox{\hsize}{!}{\includegraphics[angle=270]{lineplot.ps}}
\resizebox{\hsize}{!}{\includegraphics[angle=270]{bump.ps}}
\caption{\emph{Upper panel}: Least squares, deconvolved line profiles of 
         CoRoT-11 (points with error bars), as obtained from the time-series 
         HIRES spectra acquired during the transit of the planet. The 
         limb-darkened, rotationally-broadened line profile of the best fitting 
         model is shown with solid lines. Radial velocity (wavelength) increases 
         from left to right, time from bottom to top. The planet signature 
         is the bump that migrates from about $-20$~\kms\ near the 
         middle of the time series, to about $+15$~\kms\ in the upper line profile. 
         \emph{Lower panel}: Residuals of the composite spectral line 
         profiles of CoRoT-11 following the subtraction of the 
         above-mentioned synthetic profile. Orbital phase 
         increases vertically. The two vertical dashed lines mark the 
         $\pm$\vsini\ of the star. The three crosses denotes the first, 
         second, and third planet's contacts. A moderately bright 
         feature, ``travelling'' from the bottom left to the top right corner, 
         appears in the residual map as the result of the transit of the 
         planet across the stellar disc.}
\label{CoRoT11b_trace}
\end{center}
\end{figure}

For each exposure, a composite stellar line profile was computed using 
the least squares deconvolution (LSD) algorithm of 
\citet{Donati1997}. This is the line profile that, when convolved 
with a mask of delta functions at the wavelengths of known 
lines, with relative depths computed for an F6\,V spectrum using 
a Kurucz atmosphere model \citep{Kurucz1979}, yields an optimal 
inverse-variance-weighted fit to the observed spectral orders. 
The stellar line mask included 1140 lines in the 15 HIRES echelle 
orders spanning the wavelength range from 4016\,\AA\ to 4856\,\AA. 
The deconvolved profiles were computed on a linear velocity scale 
in the barycentric frame, with a velocity increment of 
1.5\,\kms pixel$^{-1}$ (Figure~\ref{CoRoT11b_trace}, upper panel).

The lower panel of Figure~\ref{CoRoT11b_trace} shows the time-series 
residuals of the composite spectral line profile in grey-scale form, 
following subtraction of the the limb-darkened, rotationally-broadened 
line profile of the best-fitting stellar model. The trailed spectrum 
reveals a narrow, moderately bright feature moving with a 
constant radial acceleration through the composite stellar line profile
from the bottom left to the top right corner. This feature is identifiable 
as the spectral signature of the photospheric light blocked by 
CoRoT-11b as it transits in front of the stellar disc. Important 
results can be drawn from a qualitative inspection of the plot in 
Figure~\ref{CoRoT11b_trace}. First, the planet's Doppler shadow 
appears over the approaching limb of the star and moves towards 
the receding half, confirming the prograde orbit scenario 
suggested by the shape of the observed RV anomaly \citep{Gandolfi2010}. 
The absolute value of the radial velocity of the light blocked by 
the planet at the beginning of the transit is $\sim20$~\kms, which 
agrees with the expected value for a planetary transit with an 
impact parameter $b=0.81$ and for a stellar projected rotational 
velocity \vsini\,=\,40~\kms\ (see Table~\ref{Table_Comparison}). 
At the phase of mid-transit, the trail has a zero barycentric 
velocity relative to the star, i.e., the planet crosses the 
sky-projected stellar spin axis at the mid-point of the transit. 
Taking into account the non-central planetary transit ($b\ne0$), 
this indicates that the CoRoT-11 system is closely aligned in the 
plane of the sky ($\lambda\simeq 0\degr$).

The time-series spectra were modelled together with the CoRoT 
photometry and the out-of-transit RV data from HIRES, HARPS, 
SOPHIE, and TLS/Coud\'e, as listed in \citet{Gandolfi2010}. 
The system parameters were computed using the MCMC approach 
described by \citet{Cameron2010a}, in which the photometry, 
the out-of-transit RV data, and the least squares deconvolved 
rotation profiles are fitted simultaneously.

The photometric transit model employs the algorithm of 
\citet{Mandel2002} in the small-planet approximation, using a 
four-coefficient non-linear limb-darkening model with coefficients 
interpolated from the tabulation of \citet{Claret2004} to match 
the CoRoT bandpass for a star with the same effective temperature, 
metallicity, and surface gravity as CoRoT-11. The light 
contamination fraction of $13.0\pm1.5$~\% \citep{Gandolfi2010} was 
taken into account in modelling the CoRoT photometry. The 
modelling procedure is described by \citet{Cameron2007} and 
\citet{Pollacco2008}, and yields five parameters: the orbital 
period $P$, the epoch $T_0$ of mid-transit, the planet-to-star 
area ratio $(R_p/R_\star)^2$, the scaled stellar radius 
$R_\star/a$ as a fraction of the orbit's semi-major axis $a$, 
and the dimensionless impact parameter $b=a\cos i_p/R_\star$, 
where $i_p$ is the inclination of the planet's orbital axis to 
the line of sight. The stellar mass is estimated at each time 
step using the empirical calibration of \citet{Enoch2010}, based 
on the eclipsing-binary data compilation of \citet{Torres2010}. 
The out-of-transit RV curve was fitted using a circular 
orbit model of velocity amplitude $K$. A separate zero-point 
velocity was fitted for each of the different RV data sets.

\begin{figure} 
\begin{center}
\resizebox{\hsize}{!}{\includegraphics[angle=270]{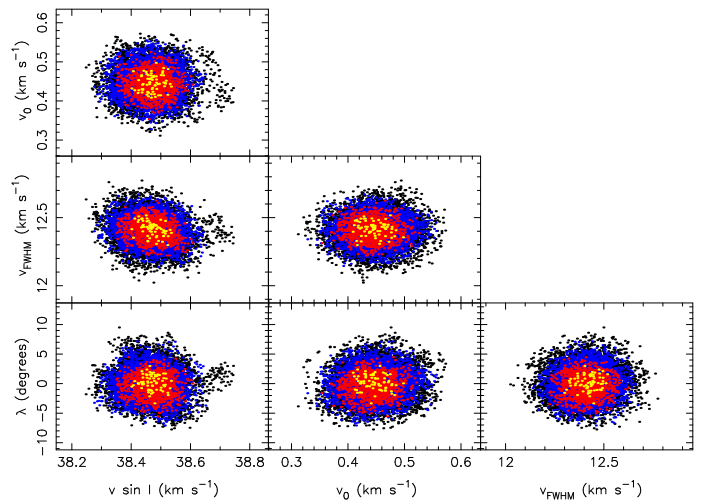}}
\caption{Correlation diagrams for the joint posterior probability 
	 distributions of the four tomography parameters, namely 
	 $\lambda$, \vsini, $v_{\rm FWHM}$, and $v_{0}$. The 
	 colours of the dots reflect the $\Delta\chi^2$ of each 
	 point, with thresholds at 2.30, 6.17, and 11.8, to show 
	 the two-parameter one, two, and three-sigma confidence 
	 regions in yellow, red, and blue, respectively (see the 
	 online edition of the journal for a color version of this 
	 figure).}
\label{CorrelationDiagrams}
\end{center}
\end{figure}

For the line-profile analysis during the transit, we used the 
line-profile decomposition model of \citet{Cameron2010a}, in which 
the stellar profile is modelled as the convolution of a narrow 
Gaussian of width $v_{\rm FWHM}$ (representing the combined 
instrumental and intrinsic stellar photospheric line profile) and 
a limb-darkened rotation profile of width \vsini. The light blocked 
by the planet was treated as a Gaussian of the same width as the 
intrinsic profile, whose strength was modulated using the model of 
\citet{Mandel2002}. The trajectory of the planet's signature in 
velocity space is governed by the projected spin-orbit misalignment 
angle $\lambda$, the stellar \vsini, the stellar radial acceleration, 
the transit impact parameter $b$, and the zero-point of the velocity 
scale on the LSD profile ($v_{0}$). The latter parameter may differ 
significantly from the systemic velocity $\gamma$ established by the 
out-of-transit RV measurements, as it depends on the form of the 
stellar spectrum and the line weights and wavelengths used in the 
LSD procedure. In all, the trailed spectrogram was modelled 
as a function of four new MCMC parameters ($\lambda$, \vsini, 
$v_{\rm FWHM}$, and $v_{0}$), in addition to the RV curve 
semi-amplitude $K$ (which is also constrained by the RV solution) 
and the impact parameter $b$ (which is also determined by the 
photometry). Figure~\ref{CorrelationDiagrams} shows the correlation 
diagrams for the joint posterior probability distributions of the 
four tomography parameters, as derived from our MCMC analysis. 
The derived system parameters and their uncertainties are listed in 
Table~\ref{Table_Comparison}.

\section{Discussion}

After \object{HD189733b} \citep{Cameron2010a}, \object{WASP-33b} 
\citep{Cameron2010b}, and \object{WASP-3b} \citep{Miller2010}, CoRoT-11b 
is the fourth transiting exoplanet for which the $\lambda$ angle has been 
derived using a Doppler imaging technique. This letter has demonstrated 
once more the advantage of this method relative to standard RV approaches 
for the analysis of the RM anomaly, especially in cases where accurate RV 
measurements are rendered unfeasible by the rapid stellar rotation rate. 
The RM effect of CoRoT-11b was first observed by \citet{Gandolfi2010}, 
who detected the RV anomaly during the transit of the planet. Although 
the shape of the RV anomaly was clearly indicative of a prograde orbit, 
the sky-projected obliquity of the system could not be constrained from the 
RV data. The measurements presented in \citet{Gandolfi2010} did not cover
the entire transit and were strongly affected by the high \vsini\ of the 
star, preventing any accurate modelling of the RV anomaly. We reanalysed 
the existing on-transit spectra with the line-profile tomography technique 
described in \citet{Cameron2010b} and confirmed a prograde orbit for 
CoRoT-11b. By tracking the trajectory of the missing starlight across 
the line profile, we measured the sky-projected system's obliquity 
$\lambda=0.1\degr\pm2.6\degr$. The advantage of tracking the Doppler
shadow of a transiting planet compared to modelling the RV anomaly 
occurring during the RM effect, is that the planet's trail is a straight 
line and its position can be extrapolated even with a partial 
spectroscopic coverage of the transit. Modelling the RV anomaly can 
lead to a degeneracy of the solution, especially if the RV data are noisy 
and do not cover the entire event.

A combined MCMC analysis of the available CoRoT photometry and 
spectroscopic data, yields the system parameters listed in 
Table~\ref{Table_Comparison}. A comparison with the previous values 
reported in the discovery paper and those published recently by 
\citet{Southworth2011}, illustrates the very good agreement (within 
$1\sigma$) of the three sets of results (Table~\ref{Table_Comparison}).

\begin{table*}[ht]
  \centering 
  \caption{CoRoT-11 system parameters as derived from our MCMC analysis 
           and comparison with the values reported in the literature.}
  \label{Table_Comparison}
\begin{tabular}{lccccl}
\hline
\hline
\noalign{\smallskip}

Parameter                                    & Symbol           & Gandolfi et al. (2010)    &   Southworth (2011)  & This work                 & Unit \\

\noalign{\smallskip}
\hline
\noalign{\smallskip}

Orbital period                               & $P$              &  $2.994330\pm0.000011$    &       --              & $2.994325\pm0.000021$     & days      \\
Transit epoch                                & $T_0$            &  $2454597.6790\pm0.0003$  &       --              & $2454690.49838\pm0.00016$ & days      \\
Transit duration                             & $d_{tr}$         &  $2.5009\pm0.0144$        &       --              & $2.44488\pm0.01800$       & hours     \\
Impact parameter                             & $b$              &  $ 0.8180\pm0.0080 $      &       --              & $0.8108\pm0.0077$         & $R_\star$ \\
Planet-to-star area ratio                    & $(R_p/R_\star)^2$&  $ 0.011449\pm0.000107$   & $0.011381\pm0.000483$ & $0.011600\pm0.000230$     &           \\
Scaled star radius                           & $R_\star/a$      &  $ 0.1451\pm0.0017 $      & $0.1452\pm0.0022$     & $0.1416\pm0.0024$         &           \\
Orbit inclination                            & $i_p$            &  $  83.17\pm0.15 $        & $83.13\pm0.19$        & $83.41\pm0.17$            & deg       \\
Orbit semi-major axis                        & $a$              &  $ 0.0436\pm0.005$        & $0.0440\pm0.0016$     & $0.04351\pm0.00036$       & AU        \\ 
Orbit eccentricity                           & $e$              &  $0$ (fixed)              & $0$ (fixed)           & $0$ (fixed)               &           \\
RV semi-amplitude                            & $K$              &  $  280.0\pm40.0 $        &       --              & $304.0\pm32.0$            & \ms       \\
Systemic radial velocity\tablefootmark{a}    & $\gamma$         &  $-1.336\pm0.044 $        &       --              & $-1.342\pm0.041$          & \kms      \\
Star mass                                    & $M_\star$        &  $   1.27\pm0.05 $        & $1.26\pm0.14$         & $1.23\pm0.03$            & \Msun     \\
Star radius                                  & $R_\star$        &  $   1.37\pm0.03 $        & $1.374\pm0.061$       & $1.33\pm0.04$            & \Rsun     \\
Star mean density                            & $\rho_\star$     &  $   0.69\pm0.02 $        & $0.69\pm0.03$         & $0.74\pm0.04$             & \gcm3     \\
Star projected rotational velocity           & \vsini           &  $    40\pm5     $        &       --              & $38.47\pm0.07$            & \kms      \\
Sky-projected spin-orbit angle               & $\lambda$        &          --               &       --              & $0.1\pm2.6$               & deg       \\
Planet mass                                  & $M_{p}$          &  $   2.33\pm0.34 $        & $2.34\pm0.39$         & $2.49\pm0.27$             & \Mjup     \\
Planet radius                                & $R_{p}$          &  $   1.43\pm0.03 $        & $1.426\pm0.057$       & $1.390\pm0.033$           & \Rjup     \\
Planet density                               & $\rho_p$         &  $ 0.99\pm0.15   $        & $1.01\pm0.16$         & $0.93\pm0.12$             & \gcm3     \\
Equilibrium Temperature                      & $T_{eq}$         &  $   1657\pm55   $        & $  1735\pm34 $        & $1715\pm36$               & K         \\                  

\noalign{\smallskip}
\hline
\end{tabular}
\tablefoot{
  \tablefoottext{a}{As derived from the HARPS/Standard RV measurements.}\\
}
\end{table*}

The RM effect allows us to measure only the sky-projected angle
$\lambda$. The true obliquity $\psi$ between the stellar spin 
axis and the orbital angular momentum can be derived only if one 
also knows the inclinations of the planetary orbit $i_p$ and stellar spin  
axis $i_\star$ with respect to the line of sight. While the transit 
light-curve modelling provides the former angle, the inclination 
$i_\star$ can be derived from the projected rotational velocity 
\vsini\ and stellar radius $R_\star$, once the rotation period of 
the star $P_{\rm rot}$ is known. Unfortunately, no evidence of 
spot-induced rotational modulation has been found in the CoRoT 
light curve, which is expected for a F6\,V star \citep{Gandolfi2010}.
The low activity level of CoRoT-11 is further confirmed by the 
lack of detectable spot-induced signatures in the composite 
stellar line profiles (cf. Figure~\ref{CoRoT11b_trace}).

\begin{figure} 
\begin{center}
\resizebox{\hsize}{!}{\includegraphics[angle=0]{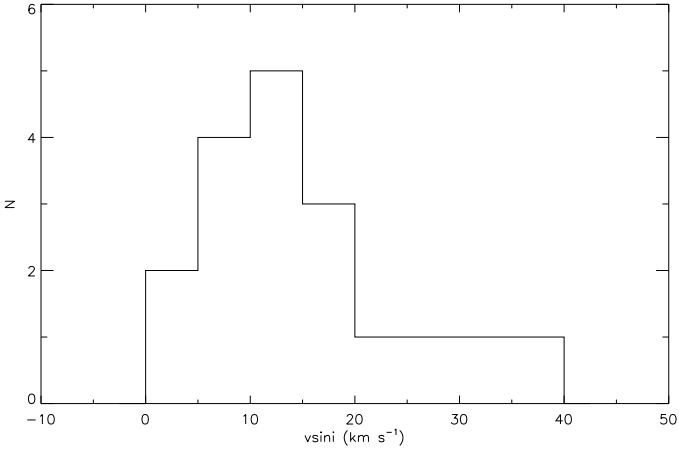}}
\caption{\vsini\ distribution of the 18 CoRoT-11-like stars from
         the compilation of \citet{Valenti2005}.}
\label{vsini_dist}
\end{center}
\end{figure}

We searched for possible evidence of an equator-on view of CoRoT-11  
by comparing its \vsini\ with the projected rotational velocity 
of a sample of randomly oriented stars similar to CoRoT-11. For
this purpose, we used the compilation of \citet{Valenti2005}, 
which provides the projected rotational velocity for 1040 field 
stars. Figure~\ref{vsini_dist} shows the \vsini\ distribution of 
the sub-sample of 18 F-type stars, whose parameters resemble those 
of CoRoT-11. Although affected by a small-number statistics, the 
histogram seems to be asymmetric and peaks between 10~\kms\ and 
15~\kms. With a projected rotational velocity of $38.47\pm0.07$~\kms, 
CoRoT-11 lies in the very right tail of the distribution, 
suggesting that $i_\star$ is likely to be close to $90\degr$.

A constraint on the true obliquity $\psi$ can be obtained from 
the variation in the duration of the transits of CoRoT-11. Since 
the star is rapidly rotating, its sizeable quadrupole moment 
induces a precession of the lines of the nodes of the orbital 
plane on a time scale as short as $\sim 7 \times 10^{4}$ years 
\citep{Lanza2011}. This produces a variation in the duration 
of the transits that can be used to constrain the obliquity. 
The effect is significant even for small obliquities 
$\psi < 10^{\circ}$ and is worth searching for in the near 
future \citep{DamianiLanza11}.

Our determination of $\lambda$ is consistent with the system 
currently being nearly aligned in the plane of the sky. This 
makes CoRoT-11 an interesting exception to the empirical rule 
stating that stars with \teff\,$\ge6250$~K and 
$M_\star\ge1.2$\,\Msun\ have high obliquities \citep{Winn2010b}. 
\citet{Lanza2011} pointed out that the tidal 
interaction time scale of the system is comparable to the 
estimated age of CoRoT-11 ($2\pm1$~Gyr) if the average 
modified tidal quality factor of the star is 
$4 \times 10^{6} \ltsim \langle Q\,_{\star}^{\prime} \rangle 
\la 2 \times 10^{7}$. One might thus wonder whether the orbit 
of the planet was initially highly tilted with respect to the 
stellar spin axis and the strong tidal torques then gradually 
aligned the system, damping down any initial misalignment. 

\begin{figure} 
\begin{center}
\resizebox{\hsize}{!}{\includegraphics[angle=0]{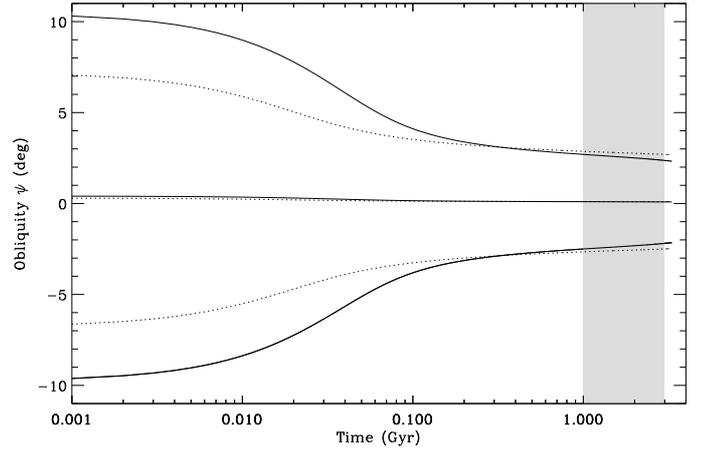}}
\caption{Forward tidal evolution of the obliquity $\psi$ for 
         $\langle Q\,_{\star}^{\prime} \rangle = 4 \times 10^{6}$ 
         (solid lines) and $\langle Q\,_{\star}^{\prime} \rangle 
         = 2 \times 10^{7}$ (dotted lines). The grey shaded area 
         indicates the age of the star (2$\pm$1~Gyr).}
\label{CoRoT11b_PsiEvol}
\end{center}
\end{figure}

To answer this question, we studied the tidal evolution of the 
obliquity, following the initial condition scenario described 
in \citet{Lanza2011}. According to this scenario, on the  
zero-age main sequence (ZAMS) the planet would have had an 
initial orbital period that was nearly synchronous with the stellar 
spin. We assumed that $i_\star\,=\,i_p\,=\,83.41\degr$ (i.e., $\psi\approx\lambda$), 
which implies a current value of the stellar rotation period 
of $P_{\rm rot}=1.74$~days. Figure~\ref{CoRoT11b_PsiEvol} shows 
the forward tidal evolution of the true obliquity $\psi$ in 
the $\langle Q\,_{\star}^{\prime} \rangle$ domain found 
by \citet{Lanza2011} for the CoRoT-11 system, namely, 
$4 \times 10^{6} \ltsim \langle Q\,_{\star}^{\prime} \rangle 
\ltsim 2 \times 10^{7}$. With a system age of $2\pm1$~Gyr, the 
current $\lambda = 0.1 \degr \pm2.6 \degr$ translates into an 
initial obliquity  $|\psi| \lesssim 10\degr$ on the ZAMS. We 
remind the reader that $\langle Q\,_{\star}^{\prime} \rangle 
\ltsim 4 \times 10^{6}$ would imply a non-synchronous initial
state of the system, as well as an age younger than 
$1$~Gyr \citep{Lanza2011}. This is in contrast to the 
estimated age (2$\pm$1~Gyr). On the other hand, if the star-planet 
tidal interaction were weak, i.e., $\langle Q\,_{\star}^{\prime} 
\rangle \ga 2 \times 10^{7}$, the system would basically not have 
evolved during its lifetime and the current low obliquity 
would be close to that of its initial state on the ZAMS. The 
possibility that $\langle Q\,_{\star}^{\prime} \rangle$ might 
be as small as $10^{6}$ in the case of obliquity evolution, while 
of the order of $10^{7}-10^{9}$ in the case of planet migration was 
suggested by \citet{Lai2012} and leads to results similar to those 
shown in Figure~\ref{CoRoT11b_PsiEvol}, pointing again towards an 
initially low obliquity. Finally, an initially high eccentric orbit 
of CoRoT-11b would change the above results. However, this circumstance 
was discussed in detail in Sects.\,3.3.1 and 4.2 of \citet{Lanza2011} 
and seems to be unlikely.

In conclusion, our results suggest a smooth, non-violent inward 
migration of CoRoT-11b during the system formation, involving 
angular momentum exchanges with the protoplanetary disc 
\citep[e.g.,][]{Lin1996}, which would not have perturbed the 
planet's primordial alignment.

\begin{acknowledgements}
We thank the editor and the anonymous referee for their careful reading, 
useful comments, and suggestions, which helped to improve and strengthen 
the manuscript. Part of the data presented herein were obtained at the W.M. 
Keck Observatory from telescope time allocated to the National 
Aeronautics and Space Administration through the agency's scientific 
partnership with the California Institute of Technology and the University 
of California. The Observatory was made possible by the generous financial 
support of the W.M. Keck Foundation. The authors wish to recognize and 
acknowledge the very significant cultural role and reverence that the 
summit of Mauna Kea has always had within the indigenous Hawaiian 
community. We are most fortunate to have the opportunity to conduct 
observations from this mountain.

\end{acknowledgements}

\end{document}